
\input harvmac
\def\square{\kern1pt\vbox{\hrule height 1.2pt\hbox{\vrule width
1.2pt\hskip 3pt \vbox{\vskip 6pt}\hskip 3pt\vrule width 0.6pt}\hrule
height 0.6pt}\kern1pt}

\def\fun#1#2{\lower3.6pt\vbox{\baselineskip0pt\lineskip.9pt
  \ialign{$\mathsurround=0pt#1\hfil##\hfil$\crcr#2\crcr\sim\crcr}}}

\def\om{\omega}

\def\p{\Phi}
\def\phi{\Phi}
\def\part{\partial}
\def\p{\phi}
\def\csi{\xi}
\def\ep{\epsilon}
\def\lta{\mathrel{\spose{\lower 3pt\hbox{$\mathchar"218$}}
     \raise 2.0pt\hbox{$\mathchar"13C$}}}
\def\gta{\mathrel{\spose{\lower 3pt\hbox{$\mathchar"218$}}
     \raise 2.0pt\hbox{$\mathchar"13E$}}}
\def\spose#1{\hbox to 0pt{#1\hss}}


\def\fun#1#2{\lower3.6pt\vbox{\baselineskip0pt\lineskip.9pt
  \ialign{$\mathsurround=0pt#1\hfil##\hfil$\crcr#2\crcr\sim\crcr}}}
\baselineskip=24truept

$\,$
\centerline{}
\centerline{\bf THE MAD ERA:}
\centerline{\bf A POSSIBLE NEW RESOLUTION TO}
\centerline{\bf THE HORIZON, FLATNESS, AND MONOPOLE PROBLEMS}
\vskip 0.2truein
\centerline{\bf Katherine Freese$^{1}$
and Janna  J. Levin$^2$ }
\vskip 0.3truein
\centerline{\it $^1$Physics Department, University of Michigan}
\centerline{\it Ann Arbor, MI 48109}
\vskip 0.15truein
\centerline{\it $^2$Physics Department, Massachusetts Institute
of Technology}
\centerline{\it Cambridge, MA 02139}
\vskip 0.4truein

\centerline{\it submitted to Physical Review Letters}
\centerline{\it November 6, 1992}

\bigskip

\medskip
\bigskip

\centerline{\bf ABSTRACT} \vskip 0.2truein

A cosmology with a dynamical Planck
mass $m_{pl}$ is shown to solve the
horizon
and monopole problems (and possibly flatness)
if there is an early MAD (modified
aging)
era where the universe becomes
older than in the standard model
as a result of a large $m_{pl}$: the
causality condition is $m_{pl}(T_c)/
m_{pl}(T_o) \gta T_c/T_o$ ($T_c$ is
some high temperature while $T_o =
2.74K$.) Unlike inflation,
there is no period of vacuum domination
nor any entropy violation. We study:
a) bare scalar theories of gravity,
b) self-interacting models, and
c) bare theories with a phase
transition in the matter sector.

 \vfill\eject

The standard Hot Big Bang model is unable to explain
the smoothness or flatness of the observed universe.
The inflationary model$^1$
solves the horizon, flatness, and monopole problems
with an era of
false vacuum domination during which
the scale factor $R$ grows
superluminally.
During inflation the temperature
of the universe drops as $T \propto R^{-1}$.
Therefore,
the next
crucial ingredient for
a successful
inflationary model is a period of
entropy violation which reheats
the universe to a high T.

We propose that a cosmology with a dynamical Planck mass can
resolve the horizon and
monopole problems without a period
of vacuum domination.
[We are still in the process of investigating
how well our model addresses the flatness problem.]
Further,
entropy production is not required.  We
have considered here general scalar
theories of gravity in which the  Planck
mass is some function of a scalar field,
$m_{pl}\propto f(\psi)$. However, we
stress that the resolution we propose
to these cosmological
problems is more generally a feature
of a dynamical Planck mass.

We consider a cosmology where the energy
density of the universe begins radiation
dominated and then goes over to a period
of matter domination as in the standard model.
In our model, in an early stage of
the radiation dominated era,
the Planck mass is very large.  Thus,
the universe is older at a given temperature
than in a standard Hot Big Bang model.
We call this epoch of `modified aging'
the
MAD era.
Larger regions of space come into
causal contact at some high T and
thereby become Smoot without violating causality.

The observable universe today (subscript $o$)
can fit inside an early causally connected
region (subscript $c$) if
        \eqn \causal{{1 \over R_c H_c} \geq {1 \over H_o R_o}\ , \ }
where $H$ is the Hubble parameter.
[Inflation satisfies this condition
by having a superluminal period
of growth of the scale factor so that
$R_o / R_c$ is very large, followed by
entropy violation].  In our
model, this condition is satisified
by having $H_c$ much smaller than
in the standard model, i.e. $t_c$ is
very large.
Extensions of Einstein gravity with a variable
Planck mass $m_{pl} = m_{pl}(t)$ can
achieve this extra aging by having a large
value of the Planck mass early on
during the MAD era.
For $H^2={(8\pi\rho/ 3 )m_{pl}^{-2}}$
(as shown below, this is the correct equation
of motion in the `slowly rolling' limit
where the time variation of the Planck mass
is sufficiently small),
the causality requirement becomes roughly
  \eqn \rough{{m_{pl}(T_c) \over M_o} \gta {T_c \over T_o}\ , \ }
where $M_o = m_{pl}(T_o) = 10^{19}$ GeV and
$T_o = 2.74$K.  For $T_c = 3 \times 10^{16}$ GeV,
e.g., this requirement becomes $m_{pl} (T_c) / M_o
\geq 10^{28}$.  We discuss below three
alternate theories of gravity which all
satisfy causality in this way.

Subsequent to $T_c$, $m_{pl}$ must move down to the value $M_o$
by the time of nucleosynthesis.
Case a) considers scalar theories of gravity
without a potential for the scalar field.
For pure Brans-Dicke gravity$^2$, the Planck mass cannot
drop to the required value in time; for models
where the Planck mass is a more complicated function
of a scalar field, our preliminary analysis indicates
that $M_o$ may be obtained.
We investigated various
additional mechanisms to drive the Planck mass down.
We considered case b), the addition of
a potential for the Planck mass,
and case c), scalar theories in the
presence of a phase transition in the matter
background.
We found that all the constraints on a potential
in case b) could not be met
without inputing small parameters into the potential,
creating a cosmological constant,
or allowing the Planck mass to be away from
the minimum of the potential today.
Case c) seems to be a viable solution to
causality which produces $M_o$
today; however, our analysis of
this case is preliminary.
Cases a) and b) have been worked out
in detail and are discussed
in two other papers$^{3,4}$.

Extended$^5$, hyperextended$^6$, and induced gravity$^7$ inflation
use modified gravity as well.  However, they differ from our
work in that they require a vacuum dominated epoch and
entropy violation, and  extended models
require an additional scalar field
as inflaton.
Since these models also need an additional mechanism,
such as a potential for the dynamical $m_{pl}$,
to drive it down
to $M_o$ by today,
the difficulties we illustrate in case b) will apply
to  these inflationary models as well.

{\bf  Action}.
In scalar theories of gravity, such
as those proposed by Brans and Dicke$^2$
and studied by Bergmann$^8$ and by Wagoner$^9$, the
Planck mass
is determined dynamically by the expectation
value of $\p$.  The action is
  \eqn\je{A=\int d^4x\sqrt{-g}\left[-{\Phi(\psi)\over 16\pi }{\cal R}
        - {\omega \over \Phi}{\partial_{\mu}\Phi
        \partial^{\mu}\Phi
        \over 16\pi} - V(\p)+{\cal L}_{\rm m}\right ]\ \ ,
        } where $\omega=8\pi{\Phi \over  (\partial\Phi/\partial
\psi)^2}$, we used the metric convention $(-,+,+,+)$, ${\cal L}_{\rm
m}$ is the Lagrangian density for
all the matter fields excluding the field
$\psi$, and $V[\p(\psi)]$ is the potential for the field $\psi$.
Stationarizing this action in a Robertson-Walker metric gives
the equations of motion for the scale factor of the universe $R(t)$ and for
$\p(t)$,
  \eqn\one{\ddot \Phi +3H\dot \Phi={8\pi(\rho-3p)\over 3+2\omega}
        -{\partial U_{ }\over \partial \Phi}
        -{\partial\omega / \partial\Phi  \over 3+2\omega}
        \dot \Phi^2 }
  \eqn\two{H^2+{\kappa \over R^2}={8\pi(\rho+V) \over 3\Phi}
        -{\dot\Phi\over\Phi} H +{\omega \over 6}\left({\dot \Phi
        \over \Phi}\right)^2 }
where
  \eqn\three{
        {\partial U_{} \over \partial \Phi}={16\pi\over
       3+2\omega}
        \left [\Phi{\partial V \over \partial \Phi}-2V\right ]; }
$U$
effectively acts as a potential term in
the equation of motion for $\p$.
$H=\dot R/R$ is the Hubble constant,
while $\rho$ is the energy density
and $p$ is the pressure in all fields
excluding the $\psi$ field.
The entropy per comoving
volume in ordinary matter, $S=(\rho +p)V/ T$, is conserved.
We define
$\bar S=R^3T^3 $ where $S \simeq \bar S (4/3)(\pi^2/30)g_*$ and
$g_*(t)$ is the number of relativistic degrees of freedom
in equilibrium at time $t$.

{\bf Case (a)  Massless Scalar Theory:  $V(\p)=0$}.
Here we consider a scalar field $\p = m_{pl}^2$ with
$V(\p)=0$. During the radiation dominated
phase, we take $\rho - 3 p =0$.
For a detailed presentation,
see Ref. (3). There we considered
two different forms of $\p (\psi)$: i) the Brans-Dicke (BD)
proposal of $\p = (2 \pi / \omega) \psi^2$ with $\omega$ constant,
and ii) general $\p (\psi)$ with $\omega$ not constant.
We found that, no matter what the
initial conditions for the BD field,
it evolves quickly towards
an asymptotic value which we call $\tilde \p = \tilde m_{pl}^2$.
At this point the equations of motion reduce
to those of an ordinary radiation dominated
Einstein cosmology with $M_o$, the usual
Planck mass of $10^{19}$ GeV, replaced by $\tilde m_{pl}$.
In particular, $R \propto \tilde m_{pl}^{-1/2} t^{1/2}$
and $H = 1/2t$.

We illustrate the derivation of these results briefly.
The equation of motion \one \ has
solution
$  \dot \p R^3=-{C (1+2\om/3)^{-1/2}}.$ $C$ is a
constant of integration and can be positive, negative, or zero. Immediately
we see that if $C\ne0$, then $\left | \dot \p \right |$,
which may initially have a large value,
shrinks as $R$ grows.
Eventually, $\p$
appears effectively constant and approaches $\tilde \p$.
If $C > 0$, then $\dot \p < 0$; $\p$ starts out
larger than $\tilde \p$ and approaches it
from above.  If $C < 0$, then $\dot \p >0$;
$\p$ starts out smaller than $\tilde \p$ and approaches
it from below.  If $C = 0$, then $\dot \p = 0$
and $\p = \tilde \p$ throughout radiation domination.

We found it easiest to parameterize $R, \ T,$ and $H$
in terms of $\p=m_{pl}^2$, and then subsequently to solve for $\p(t)$.
We present results here for the simplest
case of pure BD gravity with spatial curvature
$\kappa = 0$; more general results
can be found in Ref. (3).
Integrating eqn. \two \ , we found
        \eqn\scale{
        R(\p)={C\over 2\bar S^{2/3}\gamma^{1/2}}
        \tilde \p^{-1/2} {\rm exp}(- \Theta / 2 \epsilon)
        {1 \over {\rm sinh} \Theta} \ . \ }
Here $\gamma(t)\equiv(8\pi^3/90)g_*(t)$,
$\Theta \equiv \epsilon \ {\rm ln} ({\p / \tilde \p}) $ and
$\epsilon \equiv {\pm(1 + 2 \omega/3)^{1/2} }/2$,
where the $+(-)$ in $\epsilon$ refers to $C >(<) 0$.
{}From adiabaticity,  $T(\p)=\bar S^{1/3}/R$.
We have
$H(\p)=\dot R/R=(-C/R^4)(dR/d\p)$.
Eqns. \one \ and \two \ and the conservation of matter
equation determine
$\Phi(t)$, $\rho(t)$, and $R(t)$ up to
four constants of integration, which we can choose to be
$\bar S,\ C,\ \p(t=0), \ {\rm and}\ \  \tilde \p$.
As $\Phi$ approaches $\tilde \Phi$,
for $|\epsilon|>1/2$ ($\omega > 0)$,
$R(\p)$ grows and thus
$T(\p)$ drops adiabatically.
In addition,
the comoving horizon size grows, as does
$H^{-1}R^{-1}$.
The size of a causally connected region
can grow large enough to resolve the horizon problem.

We can write ${H_o=\alpha_o^{1/2}
{T_o^2/ M_o}}$ with
$\alpha_o=\gamma(t_o)\eta_o=8\pi/3(\pi^2/30)g_*(t_o)\eta_o$ where
$\eta_o\sim 10^{4}-10^5$.
We can use
adiabaticity, $RT=\bar S^{1/3}\propto
(S/g_*)^{1/3}$, and our solutions{} to write the causality
condition in eqn. \causal \ as
  \eqn\hee { { m_{pl}(T_c) \over T_c}
        {2\ep\over \sinh\Theta_c+2\ep\cosh\Theta_c}\gta
        \beta {M_o\over T_o}\ \ } where
$\beta=(\gamma(t_c)/\alpha_o)^{1/2}(g_*(t_c)/g_*(t_o))^{-1/3}$.
Although it is possible for the causality condition \hee \
to be satisfied while $\Phi$ is still far from $\tilde \Phi$,
we find that
 the lowest possible value of $\Phi_c^{1/2}
\propto m_{pl}(t_c)$ that solves causality is given
by $\p_c \simeq \tilde \p$.
For $\p_c\approx\tilde \Phi$
($\Theta_c \approx 0)$,
the causality condition becomes simply
${\tilde m_{pl}\over M_o}\gta \beta
{T(\tilde \p)\over T_o}$, where $m_{pl}(t_c)\approx \tilde
m_{pl}=\tilde \p^{1/2}$. We are free to
specify the temperature at which we would
like to resolve  causality
(the choice of
T at which $\p=\tilde
\p$ is equivalent to making an appropriate choice for the ratio of
arbitrary constants $\bar S/C$).
For $T_c \simeq 250$ GeV, e.g.,
condition \hee \ requires
$\tilde m_{pl} \geq 10^{13} M_o$.

We can verify that the resolution to
the horizon problem is explained by
an old universe.
When $\p\approx \tilde \p$,
the universe evolves as an ordinary
radiation dominated universe with
$M_o$ replaced by
$\tilde m_{pl}$.  We can express
the age of the universe in terms of
$t_{\rm einst}$, the standard age in a
cosmology described by Einstein gravity:
$t(\tilde \p)=t_{\rm einst}\left({\tilde m_{pl}/ M_o}\right)$
at a given temperature.
Since $\tilde m_{pl} \gg M_o$,
the universe is older than in the standard cosmology,
e.g., at $T_c = 3 \times 10^{16}$ GeV,
$t_{\rm einst}\sim 10^{-40}$ sec while
$t(\tilde \p)\geq 10^{-12}\ {\rm sec}$.

As the universe cools below
matter radiation equality,
the nature of the solutions changes.
Thus there is a built in off-switch
to end the unusual radiation dominated
behavior of $R(\p),T(\p)$ and $H(\p)$.
The obvious difficulty with this
resolution to the  horizon problem is fixing
the value of the Planck mass to be $M_o$ by
the time of nucleosynthesis.
For pure BD, observations constrain
the parameter $\omega \geq 500$.  The rate
at which $\p$ changes is very suppressed for large $\omega$.
As an extreme example, for $\omega = 500$ and
$\tilde \p^{1/2} = 100 M_o$ at $T_c \sim 1$ eV,
then today $\p_o^{1/2} \geq 80 M_o$.
For more general scalar-tensor theories
with $\omega \neq$ constant, our preliminary
analysis indicates that $M_o$ may be obtained.
Below, we discuss the possibilities of using a potential
or an appropriate background matter (or vacuum) field
to drive $m_{pl}$ to $M_o$ by nucleosynthesis.

{\bf Case (b) Self Interacting Scalar Theory: $V(\p)\ne 0$}.
A thorough treatment
can be found in Ref. (4).
We sketch here the difficulties encountered in
simultaneously matching all the constraints
on the model.
We have only considered the slowly
rolling limit where ${\dot \p / \p} \ll H_R$,
where $H_R^2 = [8 \pi (\rho + V)/ 3 \p]$,
so that $H \simeq H_R$
and the causality condition holds as
in eqn. \rough \ .  In future work,
it would be interesting to consider the
opposite limit,
where eqn. \rough \ would be modified.

There are two possible ways to satisfy the causality
condition.  First, for a scenario
in which the potential is inconspicuous during
the early radiation dominated era, the
results of case a) are recovered.
The sole purpose of the potential would be
to push $m_{pl}$ down
to $M_o$ after the causality condition was satisfied.
Alternatively, one could imagine
a potential with interactions large enough to thermalize a
bath of $\p$ particles and
drive a phase transition.
Then, ideally, the field could reside in the
high-T minimum of the potential
$\p = \tilde m_{pl}^2$ for $T>T_c$ and quickly move to
the low-T minimum of the
potential $\p = M_o^2$ for $T<T_c$.
If $\tilde m_{pl}/M_o \geq T_c/T_o$,
then the causality condition would be
satisfied.  In the cases we considered,
the high-T minimum changes
as T drops, and, unfortunately,
the field does not stay in the minimum.
However, in principle causality could
still be solved as long as $m_{pl}(T_c)
= \tilde m_{pl}$ satisfies the above condition.

In either case, the model must satisfy the following list
of constraints:
1) The cosmological constant today is below the observational
limits,
$\Lambda_o = 8 \pi V(T_o, M_o^2)
      / M_o^2 < 10^{-122}M_o^2$.
2) The Planck mass today is a minimum
of $U_{}$,
${\partial U_{}\over \partial \p} |_{M_o^2}=0$
and
${\partial^2U_{}\over \partial \p^2} |_{\p=M_o^2}
        \equiv m_{\rm eff}^2>0$.
[Note that $\p$ is driven to the minimum
of $U_{}$ (not of V) in eqn. \three \ ].
3)
For Brans-Dicke like
theories with $\p = (2 \pi / \omega) \psi^2$
and constant $\omega$,
time delay experiments require $\omega \geq 500$
if $m_{\rm eff} \leq 10^{-27} {\rm GeV}$;
these experiments place no bounds$^{10}$ if
$m_{\rm eff} \geq 10^{-27} {\rm GeV}$.
[Experimental bounds on the time variations
of G require $|{\dot G \over G}| = |{\dot \p \over \p}|
\leq H_o$ and are automatically satisfied
in the slow-roll limit].
4) The universe is radiation dominated, not potential
dominated, at high T,
$V(\p)\lta\rho_{\rm rad}$.

As an example,
we consider the  Brans-Dicke coupling $\p=(2 \pi / \omega) \psi^2$
with the potential
$V = {\lambda \over 4} \psi^4 - {m^2 \over 2} \psi^2 + \delta$.
To obtain numbers, as an example we take $T_c = 3 \times 10^{16}$ GeV.
Constraint 4) is most restrictive right
at $T = T_c$ and requires $\lambda \omega^2 \leq 10^{-119}$.
The only case that satisfies
the combination of the first three of the constraints
is $V = {\lambda \omega^2 \over 16 \pi^2}(\p - M_{o}^2)^2$.
This potential satisfies $\Lambda_o \propto V(\p = M_o^2) = 0$.
Using this potential, we can write constraint 4)
in terms of $m_{\rm eff}$ as
$(3 + 2 \omega) (m_{\rm eff}/M_o)^2 \leq 10^{-119}$,
or $m_{\rm eff} \leq 10^{-41}$ GeV;
constraint 3) can then only be satisfied
if $\omega \geq 500$.
Constraint 4) then
requires $\lambda \leq 10^{-122}$.
With such small self-coupling the
$\p$ field will not thermalize,
there are no thermal corrections
to the potential, and there is no phase
transition.
Satisfying time delay experiments as
well as all the other constraints
can only be accomplished with a
potential that is very flat (and probably
fine-tuned).

Alternatively, consider the case of $\omega \ll 1$.
We find that, in general, all
four constraints cannot be simultaneously
satisfied (even if a linear term or a $\psi^6$ term
is added to the potential).
We cannot use
the potential after $T_c$ to pin the field
in the minimum in the presence of all four
constraints.  We must relax one of them.
For example, the field could be moving
slowly somewhere away from the minimum, i.e.,
constraint 2) does not hold.  Then
reasonable values of the other parameters
($\lambda =1$) insist on a small value
of $\omega \leq 10^{-60}$ to satisfy the
constraints; i.e. gravity must deviate
substantially from Einstein gravity.
However, in this case, the potential
is not really playing its intended role
of pinning the Planck mass in its correct value today.
In addition, compatibility of
such a small value of $\omega$
with observational bounds
on $\dot G/G$ depends on exactly
which potential is chosen.
Alternatively, there may be solutions
to the cosmological constant problem
which drive $\Lambda$ to zero without affecting the
parameters in constraint 2); then
one could relax constraint 1).
Also, modifications to the model (e.g.
$d \omega / d \p
\neq 0$)
change the constraints and may allow
more freedom in the parameters.
Generically, we expect
the feature that survives will be substantial
deviation from Einstein gravity, i.e.,
a small Brans-Dicke parameter.

If we allow the universe to become
potential dominated at some point,
we can relax constraint 4),
and the parameters of the potential
do not need to be small.
It might be possible to construct a hybrid model combining
some inflation with some MAD expansion.

Inflationary scenarios which use modified
gravity, such as (hyper)-extended inflation,
also need a potential to anchor $m_{pl}$ at $M_o$
today.  There constraint 4) would be replaced
by $V(\p) < V({\rm inflaton})\sim \rho_{\rm rad}(T_c)$.  The problems
outlined in this section would directly
apply to these models as well.

{\bf Case (c) Phase Transitions in the matter sector}.
If $V(\p)=0$ then eqn. (4) becomes
$\ddot \p+ f\dot \p={8\pi (\rho-3p)/ (3+2\om)}$, where
$f$ is a friction term.  All
contributions to $\rho-3p$ will affect the dynamics
of $\p$.
Any matter or vacuum energy with
$\rho - 3 p < 0$ on the right hand side
of the equation will tend
to drive $\p$ to a smaller value;
we will try to use this to drive $\phi$
to $M_o$ after causality has been solved.
Possibilities
include
i) a decaying negative vacuum energy and ii) a thermally
corrected matter potential
at temperatures above a phase transition.
Case i) has a negative vacuum energy that decays
in time, similar to the positive decaying vacuum
energy considered previously$^{11}$.
Here we focus on case ii).

Consider the matter Lagrangian,
 ${\cal L}_{\rm m}=-(1/2)\partial_{\mu}\eta
        \partial^{\mu}\eta -V(\eta) $ where $\eta$ is a scalar
field and the bare, uncorrected potential is
$V(\eta)=(\lambda/ 4)(\eta^2-\eta^2_{\rm
min})^2$
 with $\eta_{\rm min}=m/\sqrt{\lambda}$. With
thermal corrections, for $T\gg T_{cr}$, we have
  \eqn\therm{\eqalign{ V(T,\eta) \simeq {\lambda \eta^4\over 4}&+
        {1\over 2}\left({{\lambda
               T^2\over 4}-m^2}\right )\eta^2+{m^4\over 4\lambda}
      -{\pi^2\over 90}T^4-{T^2 m^2\over 24}\ \ . \cr}}
At high T the minimum of the potential is
at $\eta_1=0$. When T falls below
$T_{cr}=2m/\sqrt{\lambda}$, then a new global minimum appears at
$\eta_2=m/\sqrt{\lambda}$.  We chose
the  potential so that $V(\eta_2)=0$.  Now,
$\rho-3p=4V-T(\part V/\part T)$.
For $T > T_{cr}$ (where $\eta = 0$), we define
${-\alpha(T)\equiv \rho-3p=-{T^2m^2/ 12}
   +{m^4/\lambda}}$. Since this is negative,
at high T the background matter potential will
work to push $m_{pl}$ to smaller values.

To illustrate we choose
$\p=\p_o+\csi\psi^2$ so
that $\om={4\pi\phi \over \csi(\p-\p_o)}$.
Eqn. (4) becomes
    \eqn\after{\ddot\p+ \left(3H+ {\dot \omega \over 2 \omega +3}\right)
      \dot\p=
      -{8\pi\over (3+2\om)}\alpha(T) \ \ .}
Imagine the scenario to proceed as
follows:  Very early on,
the Planck mass evolves as described in
case a) above. The high
T contributions to the matter
background will not strongly affect
the $\p$ evolution.
Once $\p$ approaches $\tilde \p$
(and $\dot \phi$ becomes small),
causality is solved.
Since $\dot \p$ becomes very small,
subsequently the
$\alpha(T)$ term dominates and pushes
$\Phi \rightarrow \p_o$.
Thus, the Planck mass is driven to
$m_{pl}=\p_o^{1/2}\equiv M_o$.  At $\p=\p_o$,
$\om\rightarrow\infty$ and the motion
of $\p$ is effectively turned off.
Further work on this proposed scenario
is required to see if the Planck mass
can indeed reach its present value
prior to nucleosynthesis.

{\bf The Flatness and Monopole Problems}.
If there is a Grand Unified Epoch at temperature
$T_G$, then magnetic monopoles are produced,
typically one
per horizon volume.
In the standard cosmology, far too many are produced.
If there is inflation at $T < T_G$,
then the monopoles are inflated away.
In our model of MAD Expansion, the monopole
problem can be resolved as well.
The number of monopoles in our observable universe
today is given by the number of comoving horizon
volumes at $T_G$ that would fit inside the comoving
volume of our observable universe,
$N = \left({M_o \over m_{pl}(T_G)} {T_G \over T_o}\right)^3$.
Thus, in our model, for $m_{pl}(T_G) \gg M_o$
(a requirement similar to that for causality),
the number of monopoles in our observable universe
can be very small.

The universe can become flatter in a MAD  cosmology.
In the slow-roll
limit where $\dot \p/\p$ can
be neglected,
one can
write$^6$ $\Omega = {1\over 1-x}$, where $\Omega = \rho / \rho_c$,
$\rho_c = 1.88
\times 10^{-29} h_o^2$ gm cm$^{-3}$, $h_o = H_o / 100$ km s$^{-1}$
Mpc$^{-1},$ and $x = {\kappa/R^2 \over 8 \pi G \rho /3}$.
The observations that $\Omega_o = O(1)$ would require
$\Omega (10^{-43}\ {\rm sec}) - 1 \simeq O(10^{-60})$ in the standard Hot
Big Band model.
In our model, as the universe progresses from $T_{\rm c}$
to nucleosynthesis,
$m_{pl}$ changes from $\tilde m_{pl}$ to $M_o$.
We can calculate the ratio
${x_{nuc} \over x_c} = {\p_{nuc} \over \p_c} {T_c^2 \over T_{nuc}^2}
\leq {T_o^2 \over \beta^2 T_{nuc}^2} \simeq 10^{-17}$,
where the second relation follows from eqn. \hee \ .
In the standard model, on the other
hand, $x$ would have grown by a factor $(T_c/ T_{nuc})^2$,
e.g., for $T_c = 10^{16}$ GeV, by a factor $10^{55}$.
Thus, our model assists the approach to flatness
$(\Omega \rightarrow 1)$
by causing $x$ to become many orders of magnitude
smaller than in the standard model.
However, because of the large early $m_{pl}$
and thus small Planck time, $\Omega$ would
veer away from $1$ very quickly; we
are checking to see
if this generates the
same flatness problem as the standard model,
only at higher temperatures.

{\bf Conclusion}.
In a cosmology with a large Planck mass,
the universe grows older at a given high
temperature than in a standard cosmology--
old enough to explain how one
end of our observable universe could
have communicated with the other end
if the Planck mass satisfies $m_{pl}/M_o\gta
T_c/T_o$.
This
extra aging of the universe during the MAD era is not
in conflict with the observations of the age
of the universe, which only place limits
on the time elapsed since stars
formed.
We found that scalar theories
coupled to gravity could
slow the evolution of the universe so
that the smoothness of our observable
horizon volume is predicted.
Additional mechanisms were
proposed to anchor the Planck mass
at today's value by nucleosynthesis.

\bigskip
{\bf Acknowledgements.}
We thank F. Adams, A. Guth
and H. Zaglauer for
helpful conversations.  We thank
the ITP at U.C.S.B., the Max Planck Institut
in Munich and the
Aspen Center for Physics
for hospitality.
We acknowledge support from NSF Grant PHY-92-96020, a Sloan
Foundation fellowship, and an
NSF Presidential Young Investigator award.

\vskip 1.0truein
\centerline{\bf REFERENCES}
\vskip 0.10truein

\item{[1]} A. H. Guth, {\it Phys. Rev.} D {\bf 23}, 347 (1981).

\item{[2]} C. Brans and C. H. Dicke, {\it Phys. Rev.}
{\bf 24}, 925 (1961).

\item{[3]} J. J. Levin and K. Freese, preprint.

\item{[4]} K. Freese and J. J. Levin, preprint.

\item{[5]} D. La and P. J. Steinhardt, {\it Phys. Rev. Lett.}
{\bf 376}, 62 (1989).

\item{[6]} P. J. Steinhardt and F. S. Accetta,
{\it Phys. Rev. Lett.} {\bf 64}, 2740 (1990).

\item{[7]} F. S. Accetta, D. J. Zoller, and M.S. Turner,
{\sl Phys. Rev. }{\bf D 31}, 3046 (1985).

\item{[8]} P.G. Bergmann, {\sl Int. J. Theor. Phys.}
{\bf 1}, 25 (1968).

\item{[9]} R.V. Wagoner, {\sl Phys. Rev. D} {\bf 1},
3209 (1970).

\item{[10]} H. Zaglauer, private communication.

\item{[11]} K. Freese, F. Adams, J. Frieman, and E. Mottola,
{\it Nucl. Phys. } {\bf 287}, 797 (1987).

\end